\documentclass[12pt]{article}
 \usepackage[dvips]{graphicx}
 \usepackage{amsmath}
 \usepackage{amsxtra}
\begin{document}
\begin{center}
\large{\bf {One-Dimensional Motion of Bethe-Johnson Gas}}\\
Alex Granik\\
 Department of Physics\\
 University of thePacific\\
Stockton,CA.95211\\
 E-mail: agranik@pacific.edu
\end{center}

\newpage
{\bf ABSTRACT}\\

{\small A one dimensional motion of the Bethe-Johnson gas is studied in a context of
Landau's hydrodynamical model of a nucleus-nucleon collision. The expressions for
the entropy change, representing a generalization of the previously known results,
are found. It is shown that these expressions strongly depend on an equation of
state for the baryonic matter.}

\endcenter
\section{INTRODUCTION}
The application of special relativistic gasdynamics to some problems in particle
physics goes back to the pioneering work by L.Landau \cite{LL} who obtained the
asymptotic solution describing a one-dimensional expansion of a finite slab of
matter modelled by the ultra-relativistic gas with the adiabatic index $\gamma=
4/3$. Later I.Khalatnikov \cite{IK} obtained the exact solution of the same problem.
G.Milekhin \cite{GM} extended this solution to an ultra-relativistic gas with an
arbitrary adiabatic index. S.Belenkiy and G.Milekhin \cite{SB} applied a model of
the ultra-relativistic shock wave to a problem of an entropy increase in a high
energy nucleus-nucleon collision. A significant input into the problems associated
with a hydrodynamical model of nucleus-nucleon collisions was made by G.Baym
\cite{GB} who solved the Landau problem \cite{LL} for the cylindrical geometry.
Nonetheless G.Baym and collaborators indicated that there is a need to use a more
realistic equation of state than the one with $\Gamma=4/3$. Since there is still
some uncertainty about an exact form of the equation of state for the baryonic
matter at very high densities, one can consider a wide range of possible candidates
for such an equation.\

One of these equations was proposed by H.Bethe and M.Johnson \cite{HB} (which in
what follows we call $BJ$ equation), and later it was used to  illustrate an idea of
a possible baryon-quark transition \cite{GC}. This equation was also used for a
speculation on a potential influence of such a transition on a collective flow in a
high energy ion collisions \cite{AG}. Therefore, it seems reasonable that a study of
one-dimensional motions of Bethe-Johnson gas( including the original Landau problem)
might provide a new insight into high energy phenomena within a model based on
relativistic gasdynamics.
\section{SHOCK ADIABATICS OF BJ GAS}
We begin by considering a relativistic shock. In what follows we adopt the units
with the speed of light taken to be $c=1.$ The energy, momentum, and particle
conservation laws across the shock (in the shock frame)yield the following shock
adiabatic \cite{AT}
\begin{equation}
\label{1} \mu_1-\mu_2+\frac{1}{2}(p_1-p_2)(\hat{V}_1+\hat{V}_2)=0
\end{equation}
where indices $1$ and $2$ denote the areas in front and the back of the shock
respectively, $\mu\equiv W^2/2$, $W=w/n,~~~w=\epsilon+p$ is the enthalpy  per  unit
of  volume per unit of  baryon rest mass, $n$ is the particle density, $p$ is the
pressure per unit of baryon rest mass, $\epsilon$ is the energy density per unit of
baryon rest mass,
\begin{equation}
\label{A0} \hat{V}=W/n=w/n^2
\end{equation}
is the generalized "specific volume" \cite{KT}. In these variables the relativistic
shock adiabatic (\ref{1}) looks exactly as the
Hugoniot adiabatic of a newtonian gas.\\

\subsection{SOME RELATIONS FOR A RELATIVISTIC SHOCK }

We provide a series of relations which follow from the conservation laws across the
shock. From the continuity of particle flux density  $$ J\equiv n_1U_1=n_2U_2$$ and
momentum flux density $$(\frac{W}{n})_1U_1^2+p_1=(\frac{W}{n})_2U_2^2+p_2$$ (where
we use the definition $W=w/n$) we obtain
\begin{equation}
\label{insert0} J^2=\frac{p_2-p_1}{\hat{V}_1-\hat{V}_2}
\end{equation}
 indicating that either
\begin{eqnarray}
\label{insert} a) p_2>p_1,~~~ \hat{V}_1>\hat{V}_2\nonumber\\
b)p_2<p_1,~~~ \hat{V}_1<\hat{V}_2
\end{eqnarray}
in exact analogy to the newtonian case.It has turned out that only case $a)$
 can actually occur.\\

This is easily shown by considering  a relativistic weak shock wave, in which the
discontinuity in every quantity is small as compared to the quantity. Let us expand
both sides of the shock adiabatic (\ref{1}) in powers of small differences of
entropy(per unit of volume) $\Delta s\equiv s_2-s_1$ and pressure $\Delta p\equiv
p_2-p_1$. Following the analogous procedure in newtonian gasdynamics, the expansion
in $\Delta p$ is to be carried as far as the third order. Thus we obtain:
\begin{eqnarray}
\label{i1} \mu_1-\mu_2=W_1^2-W_2^2=(\frac{\partial W^2}{\partial s})_1\Delta
s+(\frac{\partial W^2}{\partial p})_1\Delta\
p+\nonumber\\(\frac{1}{2}\frac{\partial^2 W^2}{\partial p^2})_1(\Delta p)^2+
\frac{1}{6}(\frac{\partial^3 W^2}{\partial p^3})_1(\Delta p)^3
\end{eqnarray}

where according to the thermodynamic identities, $$(\frac{\partial W}{\partial
s})_p=T~(the~absolute~ ~temperature),~~~~~~~~~~~~~~(\frac{\partial W}{\partial
p})_s=V$$ Therefore (\ref{i1}) yields:
\begin{eqnarray}\
\label{i2} \mu_1-\mu_2=2W_1T_1\Delta s+2W_1V_1\Delta p+[V_1^2+W_1(\frac{\partial
V}{\partial p_1})_s](\Delta
p)^2+\nonumber \\
\lbrack V_1(\frac{\partial V}{\partial p_1})_s+\frac{W_1}{3}(\frac{\partial^2
V}{\partial p_1^2})_s\rbrack(\Delta p)^3
\end{eqnarray}
The generalized "specific volume" $\hat{V}=WV$ needs to be expanded only in terms of
$\Delta p$, since the terms with $\Delta s$ contribute to the fourth order of
smallness in $\Delta p$. Thus we get after some algebra:
\begin{equation}
\label{i3} \hat{V}_2-\hat{V}_1=W_2V_2-W_1V_1=[V_1^2+W_1(\frac{\partial V}{\partial
p_1})_s](\Delta p)^2(\Delta p)+\frac{1}{2}[3V(\frac{\partial V}{\partial
p_1})_s]\Delta p^2
\end{equation}
where once again we use the thermodynamic identity $V=(\partial W/\partial p)_s$.
Substituting this expansion together with (\ref{i2}) in (\ref{1}). we obtain
\begin{equation}
\label{i4} \Delta s=s_2-s_1= \frac{1}{12T_1W_1}(\frac{\partial^2 \hat{V}}{\partial
p_1^2})_s(p_2-p_1)^3
\end{equation}
which once gain represents the exact analogy of the respective relation in newtonian
gasdynamics (e.g., Ref. \cite {LL}). In all cases that have been studied the second
derivative
$$(\frac{\partial^2 \hat{V}}{\partial p_1^2})_s>0 $$ Therefore by
the law of increase of entropy $s_2 > s_1$  and according to (\ref{i4}) $p_2
>p_1$, which implies (Eq.\ref{insert}) $\hat{V}_1>\hat{V}_2.$  \\

It was also  shown \cite{AT} that a weak relativistic shock waves propagates with a
speed of sound $c_s^2=(\partial p/\partial e)_s.$ In fact, for small $p_2-p_1$ and
$\hat{V}_2-\hat{V}_1$ relation (\ref{insert0}) can be written, in the first
approximation, as
$$J^2=-(\frac{\partial p}{\partial\hat{V}})_s$$
In the same approximation
\begin{equation}
\label{insert3}
 U_1^2=U_2^2=U^2=J^2V^2=-V^2(\frac{\partial
p}{\partial\hat{V}})_s
\end{equation}

On the other hand,
$$(\frac{\partial p}{\partial\hat{V}})_s=(\frac{\partial
p}{\partial e})_s\frac{1}{[(\partial\hat{V}/\partial e)_s}$$
 where by using the identity $$-p(\frac{\partial V}{\partial e})_s
 =[\frac{\partial(eV)}{\partial e}]_s=e(\frac{\partial V}{\partial e})_s+V$$
 we find
\begin{equation}
\label{A1}
 (\frac{\partial V}{\partial
e})_s=-\frac{1}{n^2}(\frac{\partial n}{\partial e})_s=-\frac{V}{w}=-\frac{1}{nw}
\end{equation}
 and
\begin{eqnarray*}
(\partial\hat{V}/\partial e)_s\equiv(\frac{\partial wV^2}{\partial e})s\equiv
V^2\lbrack\frac{\partial(e+p)}{\partial e}\rbrack_s+2Vw(\frac{\partial V}{\partial
e})_s=
\\V^2(1+c_s^2)-2V^2=-V^2(1-c_s^2)
\end{eqnarray*}
As a result Eq.(\ref{insert3}) yields
$$U^2=-V^2(\frac{\partial
p}{\partial\hat{V}})_s =\frac{c_s^2}{1-c_s^2}\equiv U_s^2$$
\subsection{SHOCK WAVE IN BJ GAS}
We consider a  shock wave  propagating in a specific gas with the following equation
of state (Bethe-Johnson gas)
\begin{equation}
\label{2} p=(\nu-1)(\epsilon-n)
\end{equation}
where  $\nu$ is an arbitrary constant playing the role of the adiabatic index. If
the gas in front of the shock is cold, that is $p_1=0$ , equation (\ref{2}) yields
for that region:
\begin{equation}
\label{3} \epsilon_1=n_1
\end{equation}
which implies $\mu_1=1$ and hence $\hat{V}_1=V_1$, that is the
conventional specific volume.\\
\begin{figure}
\begin{center}
\includegraphics[width=6cm, height=6cm]{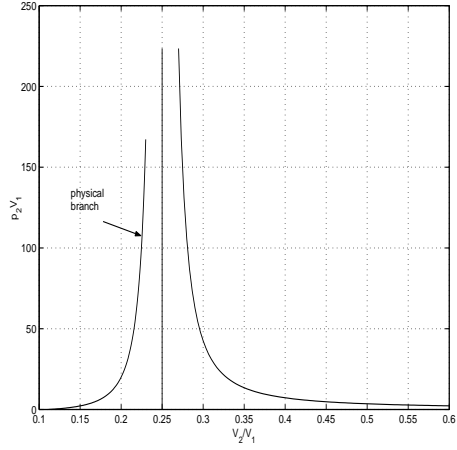}
\caption{\small Pressure behind the shock $p_2$ as a function of $\hat{V}_2/V_1$ for
$\nu=1.3$ and a cold gas ($p_1=0$) in front of the shock}
\end{center}
\end{figure}
By substituting Eqs.(\ref{2}), (\ref{3}) in the shock adiabatic (\ref{1}) we obtain
the following equation with respect to $p_2$
\begin{equation}
\label{4}
 p_2[p_2V_1(\frac{\hat{V}_2}{V_1}\frac{1}{\nu-1}-1)^2-
\frac{\hat{V}_2}{V_1}\frac{1}{\nu-1})(\nu+1)^2+1]=0
\end{equation}

 It has two solutions , one corresponding to a trivial case of a
continuous flow with $p_2=p_1=0$ ( weak shock in this case), and
the second one representing the strong shock wave:
\begin{equation}
\label{5} p_2=(\nu-1)\frac{\hat{V}_2(\nu+1)-V_1(\nu-1)}{[\hat{V}_2-V_1(\nu-1)]^2}
\end{equation}
A graph of $p_2V_1$ vs. $\hat{V}_2/V_1$ for a specific value of  the "adiabatic
index"
$\nu=1.3$ is shown in Fig.$1$.\\

From equation (\ref{5}) follows that it has two branches ( as is clearly seen in the
Fig.1), one for $\hat{V}_2/V_1\leq\nu-1$ and another one for
$\hat{V}_2/V_1\geq\nu-1.$ We cannot determine the physical branch on the basis of
the adiabatic only , since for both branches the  conditions $p_2>p_1 , V_2<V_1$
[which follow from the law of entropy increase (\ref{i4}) and Eq.(\ref{2})] are
satisfied in the region $(\nu-1)/(\nu+1)\leq\hat{V}_2/V_1\le 1$. To choose the
appropriate branch we find the adiabatic speed of sound in the region behind the
shock, $(c_s)_2$ ( it is obvious that in the
region $1$ of the cold gas the speed of sound is $0$).\\

From (\ref{2}) follows
\begin{equation}
\label{5a} (c_s)_2^2=(\nu-1)[1-(\frac{\partial n_2}{\partial
e})_s]=(\nu-1)(1-n_2V_2)
\end{equation}
where we use (\ref{A1}). We express this speed of sound in terms of $\hat{V}_2$.
From (\ref{A0}) and
$$w=\frac{\nu}{\nu-1}p_2+\frac{1}{V}_2$$ [ which follows from
(\ref{2})] we find
$$\hat{V}_2\equiv w_2V_2^2=\frac{\nu}{\nu-1}p_2V_2^2+V_2.$$ Solving this
equation with respect to $V_2$ we obtain
\begin{equation}
\label{5c} V_2=(\nu-1)(\frac{\sqrt{1+4\frac{\nu}{\nu-1}p_2\hat{V}_2}-1}{2\nu p_2})
\end{equation}
Therefore
$$w_2V_2=\frac{\hat{V}_2}{V_2}=\frac{2\nu}{\nu-1}\frac{p_2\hat{V}_2}
{\sqrt{1+4\frac{\nu}{\nu-1}p_2\hat{V}_2-1}}$$ Inserting this in (\ref{5a}) and using
(\ref{5}) we get
\begin{equation}
\label{5d} (c_s)_2^2= (\nu-1)\frac{(\nu-1)-\hat{V}_2/V_1}{\nu\hat{V}_2/V_1}
\end{equation}
Now it becomes clear that  for realistic values of $\nu\ge 1$
$$\frac{\hat{V}_2}{V_1}\le (\nu-1)$$ (that is $(c_s)_2^2\ge 0$) chooses the physical
branch of the adiabatic (\ref{5}), that is
only the left part of the Fig.1 has a physical meaning.\\

We also demonstrate that the value of parameter $\nu$ cannot exceed $2$, since
otherwise the flow becomes non-causal. Actually, from the boundaries
$$\frac{\nu-1}{\nu+1}\le\frac{\hat{V}_2}{V_1}\le(\nu-1)$$
(where the left inequality has been found above) of the physical region and relation
(\ref{5d}) follows that  the causality condition $(c_s)_2^2\le 1$ imposes  the
following restrictions on $\nu$
\begin{equation}
\label{5e} (\nu-1)^2\le\frac{\nu-1}{\nu+1}(2\nu-1),~~~~~~~~that~ is~1\le\nu\le 2
\end{equation}
\subsection{RATIO OF VELOCITIES BEHIND AND IN FRONT OF THE SHOCK}
Here we derive equation for $\chi=\beta_2/\beta_1$ which is derived from the
condition of continuity of the particle flux density
$$\frac{U_2}{U_1}=\frac{V_2}{V_1}.$$ Using (\ref{5c}) and the
adiabatic (\ref{5}) we obtain for its physical branch the following expression:

 \begin{equation}
\label{5f} \frac{U_2}{U_1}=\frac{(\nu-1)-\frac{\hat{V}_2}{V_1}}{\nu}
\end{equation}

We also provide  the expression for $\chi=\beta_2/\beta_1$ found in
\cite{AG},\cite{GC1}:
\begin{equation}
\label{6} U_1^2\chi^2-\nu U_1^2\chi+(\nu-1)(1+U_1^2-\sqrt{1+U_1^2(1-\chi^2)}=0
\end{equation}
Its implicit solution is:
\begin{equation}
\label{7} \nu= 1+U_1^2\chi\frac{\chi-1}{U_1^2(\chi-1)+\sqrt{1+U_1^2(1-\chi^2})}
\end{equation}
The respective graph of $\chi=f(\nu,U_1^2)$ is shown in Fig.2.

\begin{figure}
\begin{center}
\includegraphics[width=6cm, height=6cm]{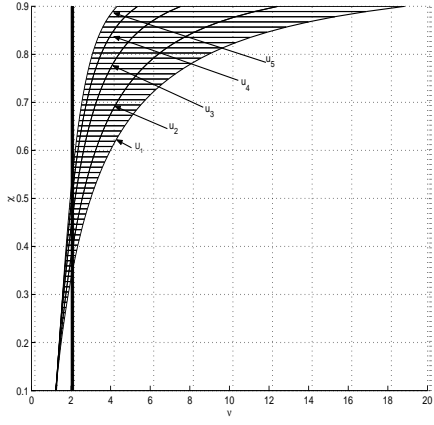}
\caption{\small {Ratio $\chi=\beta_2/\beta_1$ as a function of
$\nu$ for $5$ values of the four-velocity $U_1$ from $.1$ to $3$}}
\end{center}
\end{figure}
As is seen from the figure, for $5$ values of $\nu\le 2$ ( which
ensure the causality of the flow) parameter $\chi$ is always less
than $\nu-1$. \\

For two asymptotic cases of\

$i)$ a non-relativistic flow ($U_1\rightarrow \beta_1^2<< 1$) and\

$ii)$the ultra-relativistic flow ( with $U_1>>1$)\

expression (\ref{7}) is significantly simplified.\\

i)$U_1\rightarrow \beta_1<<1, \epsilon \rightarrow n,~~~\hat{V}_2\rightarrow V_2$.
This allows us to rewrite (\ref{5f}) as follows
\begin{equation}
\label{8a}
\frac{U_2}{U_1}=\chi[1-\frac{\beta_1^2}{2}(1-\chi^2)]+O(\beta_1^4)=\frac{\nu-1-\chi}{\nu}
\end{equation}
By representing $$\chi=\chi_0+\chi_1\beta_1^2$$ we obtain from (\ref{8a}) the final
expression for $\chi$
\begin{equation}
\label{8} \chi=\frac{\nu-1}{\nu+1}[1+\frac{2\nu^2}{(\nu+1)^3}\beta_1^2]+O(\beta_1^4)
\end{equation}
Using (\ref{8}) and the fact that $\hat{V}_2/V_1\rightarrow V_2/V_2=\chi$ in the
shock adiabatic (\ref{5}) we obtain:
$$p_2V_1=\frac{2}{\nu+1}v_1^2$$
that is the result obtained in newtonian gasdynamics,(e.g., \cite{LL}). for a strong
shock wave. This means that the lower part of the shock adiabatic ( close to
$\hat{V}_2/V_1=(\nu-1)/(\nu+1)$, which seemingly contradicts the continuity of the
particle density flux, actually represents the non-relativistic limit of the strong
shock wave in Bethe-Johnson
gas.\\

ii)$U_1>>1 (\beta_1 \rightarrow 1)$. Solving Eq.(\ref{6})we obtain
\begin{equation}
\label{9}
\chi\equiv\frac{\beta_2}{\beta_1}=(\nu-1)(1-\frac{1}{U_1}\sqrt{\frac{\nu}{2-\nu}})+O(1/U_1^3)
\end{equation}
which means that the velocity behind the shock
$$\beta_2\approx \nu-1$$On the other hand,
\begin{equation}
\label{10}
\frac{U_2}{U_1}\equiv\frac{\chi}{\sqrt{1+U_1^2(1-\chi^2)}}=\frac{1}{U_1}\frac{\chi}{\sqrt{1-\chi^2}}+O(1/U_1^3)=
\frac{(\nu-1)-\hat{V}_2/V_1}{\nu}
\end{equation}
Using (\ref{9}) in (\ref{10}) we find with the same accuracy
\begin{equation}
\label{11}
\frac{\hat{V}_2}{V_1}=\chi=(\nu-1)(1-\frac{1}{U_1}\sqrt{\frac{\nu}{2-\nu}})+O(1/U_1^2)
\end{equation}
Upon substitution of (\ref{11}) in shock adiabatic (\ref{5}) we arrive at the
following expression:
\begin{eqnarray}
\label{12} \frac{p_2}{n_2}\approx(1-\frac{\nu}{2})\frac{\nu-1}{1-\beta_2/(\nu-1)}\gg
1\nonumber\\and \nonumber\\\frac{p_2}{n_1}=\frac{p_2}{e_1}=(2-\nu)U_1^2\gg1
\end{eqnarray}
It follows then that in this case $$e_2\gg e_1,$$ which in particular means that
(\ref{12}) is true even the particle number density is not conserved. The latter is
easily verified by direct substitution of equations of state (\ref{2}) and (\ref{3})
in the
laws of energy and momentum conservation.\\

It was suggested in \cite{MN} that for a compression ratio $n_2/n_1=5 - 20$ the
baryon matter can be transformed into the quark gas. By using these values of
$n_2/n_1$ in the shock adiabatic (\ref{5}) we can find the range of the respective
transition pressures $p_2$.
 By substituting
$$\frac{\hat{V}_2}{V_1}=\frac{\nu}{\nu-1}p_2\frac{n_1}{n_2^2}+\frac{n_1}{n_2},$$
which follows from the definition of $\hat{V}_2$ and the equation of state
(\ref{2}), we rewrite (\ref{5}) as a function of $(n_2/n_2,\nu):$
\begin{eqnarray}
\label{13}
\nu^2z^3+z^2x[1-x(\nu-1)]+2(\nu-1)x^2\{[1-x(\nu-1)]^2-1-\nu^2\}\nonumber\\
+x^3(\nu-1)^3[(\nu-1)x-\nu-1]=0
\end{eqnarray}
where $z=p_2n_1$ and $x=n_2/n_1$. If we fix the value of $e_1=n_1$ as $133 MeV/fm^3$
( based on the assumption that the ratio of the bag constant $B$ entering the bag
model equation of state and $e_1$ is $0.45$ \cite{KT} and $B\approx 60 Mev/fm^3$,
then numerical solutions of (\ref{13})in terms of $p_2$ in $10^{35}dn/cm^2$ for the
following values of the parameters
$$x=5,~8,~10; ~~~~~~~~~\nu=1.7,~1.8,~1.9$$
are given in Table 1\\

\begin{table}
\begin{center}
{\bfseries Table I. Transition pressure $p_2$ for various values
of $x=n_2/n_1$ and $\nu$}\\[1.5ex]\end{center}
\begin{center}
\begin{tabular}{|c||c|c|c|}
\hline
x =$n_2/n_1$  &  5  & 8  & 10\\
\hline
$\nu=1.7$ & $3.51$ &$42.37$ &$37.88$\\
\hline
$\nu=1.8$ &$5.68$ &$27.31$ &$49.30$\\
\hline
$\nu=1.9$ &$10.67$ &$34.76$ &$42.66$\\
\hline
\end{tabular}
\end{center}
\end{table}

As is seen from the table, the transition pressure for $n_2/n_1=8$ and $10$ are of
the same order of magnitude as the ones found in
Ref. \cite{SB1}. \\

Now we consider the shock wave in an observer's  frame of reference such that the
shock velocity is $-D$, the velocities in front and the back of the shock are
$\beta_1$ and $\beta_2$ respectively. In the shock frame of references these
velocities $\beta_{1,2}'$ become
\begin{equation}
\label{14} \beta_{1,2}'=\frac{\beta_{1,2}+D}{1+\beta_{1.2}D}
\end{equation}
From the conservation of energy and momentum across the shock follows that \cite{AT}

\begin{equation}
\label{15} \beta_1'\beta_2'=\frac{p_2-p_1}{e_2-e_1}
\end{equation}
Using equations of state for Bethe-Johnson gas (\ref{2}) we obtain that for a strong
shock ($U_1>>1$):
\begin{equation}
\label{16}
\beta_2'\beta_1'=(\nu-1)(1-\frac{1}{U_1}\sqrt{\frac{\nu}{2-\nu}})+O(U_1^{-2})\rightarrow
\nu-1
\end{equation}
Using (\ref{14}) in (\ref{16}) we get a quadratic equation with respect to $D$. The
physical solution to this equation in terms of the velocities $\beta_1,\beta_2$ in
the observers frame of reference is
\begin{eqnarray}
\label{17} D=\frac{\sqrt{(2-\nu)^2(\beta_1+\beta_2)^2+
4[(\nu-1)-\beta_2\beta_1][1-\beta_2\beta_1(\nu-1)]}}{2[1-\beta_1\beta_2(\nu-1)]}\nonumber\\
-\frac{(2-\nu)(\beta_2+\beta_1)}{2[1-\beta_1\beta_2(\nu-1)]}
\end{eqnarray}
If $\nu\rightarrow 2$ then according to (\ref{17}) $D\rightarrow 1.$ The respective
graph of $D=f(\beta_1,\beta_2)$ for $\nu=1.3$ is shown in Fig. $3$.
\begin{figure}[h]
\begin{center}
\includegraphics[width=6cm, height=6cm]{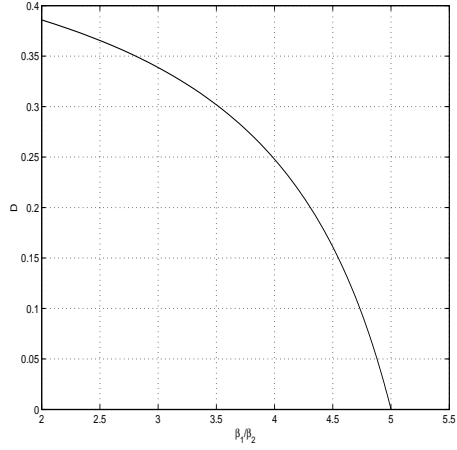}
\caption{\small {Shock velocity $D$ in the observer frame as a
function of the ratio of velocities $\beta_1$ and $\beta_2$ in
front and behind the shock respectively}}
\end{center}
\end{figure}
In what follows we need to have  the ratio $W_2(\beta_2)/W_2(0)$ where
$W_2(\beta_2)$ is the value of $W_2$ behind the shock (in the observer frame) and
$W_2(0)$ is the respective value for $\beta_2=0$. Since in a strong shock $e>>
n$,this implies $$W_2\approx\frac{\nu}{\nu-1}p_2n_2$$ Therefore we get from
(\ref{12})
\begin{equation}
\label{17a} \frac{W_2(\beta_2)}{W_2(0)}=\frac{[1-\beta_2'/(\nu-1)]_{\beta_2=0}}
{[1-\beta_2'/(\nu-1)]_{\beta_2}}
\end{equation}
Since for a very strong shock wave $\beta_1\rightarrow 1$ the velocity $\beta_1'$
also tends to $1$ according to (\ref{14}). Therefore from (\ref{16}) follows that
$\beta_2'\rightarrow \nu-1$. As a result, (\ref{17a}) becomes the indeterminacy of
the type $0/0$. Using $l'Hopitale $ rule we find (cf.\cite{SB1})
\begin{equation}
[\label{17b} \frac{W_2(\beta_2)}{W_2(0)}]_{\beta_1\rightarrow
1}=[\frac{(\partial\beta_2'/\partial\beta_1)_{\beta2=0}}
{(\partial\beta_2'/\partial\beta_1)_{\beta2}}]_{\beta_1\rightarrow 1}
\end{equation}
where $$\partial\beta_2'/\partial\beta_1=[\frac{\partial\beta_2'}{\partial
D}\frac{\partial D}{\partial\beta_2'}]$$ Performing elementary calculations we find
\begin{equation}
\label{18a} [\frac{W_2(\beta_2)}{W_2(0)}]_{\beta_1\rightarrow
1}=\frac{1-\beta_2}{1+\beta_2}
\end{equation}
\section{ONE-DIMENSIONAL SHOCK WAVE PROPAGATION}
We consider the following problem \cite{GM}: $2$ shock waves propagate in opposite
direction in a "gas cylinder" as shown in Fig.$4$. \\
\begin{figure}[h]
\begin{center}
\includegraphics[width=6cm, height=6cm]{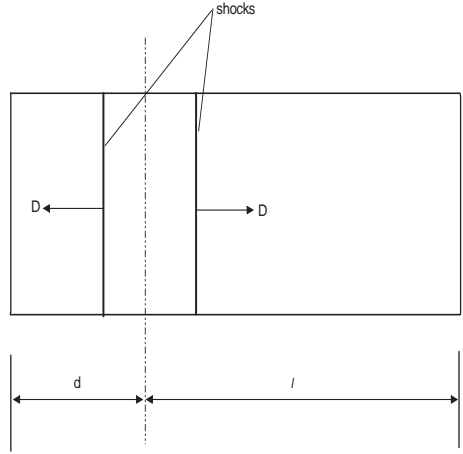}
\caption{\small {Geometry of $1$-dimensional shock waves propagation in the observer
frame of reference}}
\end{center}
\end{figure}

The distance travelled by the left-bound shock is denoted by $d$ and the respective
distance travelled by the right-bound shock is denoted by $l$.  A gas in front of
both shocks is  assumed to be "cold", that is its pressure is $p=0$. A compressed
gas is taken to be Bethe-Johnson gas, obeying the equation of state (\ref{2}).
Initially we choose the frame of reference where the gas between $2$ shocks is at
rest.\\

When the shock wave propagating to the left with a velocity $D$ reaches the left end
of the cylinder, there begins an outflow of matter. As a result,a rarefaction wave
has been formed which starts to move to the right with the speed of sound. At this
 time the shock wave moving to the right with a velocity $D$ has not yet reached the
right end. Therefore the two possibilities can materialize:\\

a)the rarefaction wave can catch up and overtake the right-moving shock wave before
it would reach the right end the right end of the cylinder,\

or\

 b)the rarefaction wave will lag behind the right-moving shock wave.\\

If we use the shock frame of reference ( of the right-bound shock), then the gas
velocity behind it is $D$, and in front of it is given by (\ref{14}). We consider a
strong shock, that is $\beta_1\rightarrow 1$ which implies that the respective
velocity in the shock frame of reference is also $\beta_1'\rightarrow 1$. Therefore
combining the relation which is obtained by dividing energy and momentum fluxes
across the shock ( see \cite{SB1}) $$\beta_1'=\frac{p_2+D^2e_2}{D(p_2+e_2)}$$ and
the equation of state (\ref{2}) we obtain the quadratic equation for $D$ whose
solution ( excluding the trivial root $D=1$) is
\begin{equation}
\label{18} D=(\nu-1)\frac{p_2/n_2}{p_2/n_2+\nu-1}
\end{equation}
Upon substitution of expression (\ref{12}) for
$$p_2/n_2\rightarrow(1-\frac{\nu}{2})\frac{\nu-1}{1-\beta_2/(\nu-1)}\gg 1 $$ ( valid
for a very strong shock with $\beta_1\rightarrow 1$ and $\beta_2\rightarrow \nu-1$)
in (\ref{18a}) we obtain that
$$D\rightarrow \nu-1$$
This means that we can find the minimum length $l_{min}$ of the "tunnel" such that
the rarefaction wave overtakes the shock wave \cite{SB1}
\begin{equation}
\label{19} \frac{l_{min}+d}{(\beta_s)_2}=\frac{l_{min}-d}{D}
\end{equation}
where the speed of sound in gas behind the shock $(\beta_s)_2$ is given by the
following expression \cite{AT}
\begin{equation}
\label{20} (\beta_s)_2^2=(\nu-1)\frac{\nu p_2/n_2}{(\nu-1)+\nu p_2/n_2}
\end{equation}
Taking into account that $p_2/n_2\gg 1$, we see that $(\beta_s)_2\rightarrow
\sqrt{\nu-1}$. Therefore from equation (\ref{19}) we get
\begin{equation}
\label{21} \frac{l_{min}}{d}=\frac{\sqrt{\nu-1}+1}{1-\sqrt{\nu-1}}
\end{equation}
For the values of $\nu=1.3-1.9$ the respective $l_{min}/d$ varies from $3.42$ to
$37.97$. In general $l_{min}/d i$ is a monotonically increasing function which tends
to infinity when $\nu\rightarrow 2$. As was shown in Refs.\cite{GM} and \cite{SB1}
the entropy change in the case of $l\le l_{min}$ is $$\Delta s \sim \frac{l}{d}+1$$

The problem becomes more involved if $l/d\ge l_{min}$, because the rarefaction wave
can overtake the shock wave but cannot pass through it. Therefore the former is
reflected from the latter thus creating the flow region bounded on one side by the
shock wave moving to the right and on the other side by the reflected wave moving to
the left. The resulting flow can be described with the help of the equation for a
general one-dimensional motion of a relativistic gas \cite{LL}
\begin{equation}\label{22}
\beta_s^2\frac{\partial^2\chi}{\partial
y^2}+(1-\beta_s^2)\frac{\partial\chi}{\partial
y}-\frac{\partial^2\chi}{\partial\eta^2}=0
\end{equation}
where $$y=Ln(W/W_0),$$ $W_0$ is the value of $W$ behind the shock when the velocity
$\beta_2=0$,

\begin{equation}
\label{22a} \eta=tanh^{-1}\beta
\end{equation}

and the generalized potential $\chi$ is connected to the spatial $x$ and temporal
$t$ coordinates as follows
\begin{eqnarray}
\label{23} x=e^{-y}(\frac{\partial\chi}{\partial
y}sinh\eta-\frac{\partial\chi}{\partial\eta}cosh\eta)\nonumber\\
t=e^{-y}(\frac{\partial\chi}{\partial
y}cosh\eta-\frac{\partial\chi}{\partial\eta}sinh\eta)
\end{eqnarray}
By using (\ref{20}) in (\ref{22}) we obtain
\begin{equation}
\label{add} (\nu-1)(1-\frac{e^{-y}}{W_0})\frac{\partial^2\chi}{\partial
y^2}+(2-\nu+\frac{\nu-1}{W_0}e^{-y})\frac{\partial\chi}{\partial
y}-\frac{\partial^2\chi}{\partial\eta^2}
\end{equation}

For a strong shock wave $\beta_s^2\rightarrow \nu-1$, $W_0\gg1, y>>1$, and equation
(\ref{add}) becomes
\begin{equation}
\label{24} (\nu-1)\frac{\partial^2\chi}{\partial
y^2}+(2-\nu)\frac{\partial\chi}{\partial y}-\frac{\partial^2\chi}{\partial\eta ^2}=0
\end{equation}
The respective boundary conditions are based on the fact that in the strong shock
wave ( in the shock frame of reference) the velocity of gas behind the shock is
$\beta_2'=\nu-1$ (cf.\cite{SB1}). Therefore from equation (\ref{14}) follows that
the shock velocity $D$ in terms of $\beta_2'$ ( in the shock frame) and
$\beta_2=-tanh\eta$ (in the observer frame) is
\begin{equation}
\label{25} D=\frac{dx}{dt}=\frac{\nu-1+tanh \eta}{1+(\nu-1)tanh\eta}
\end{equation}\

On the other hand, from (\ref{18a}) and (\ref{22a}) follows that
at the shock
\begin{equation}
\label{26} y=2\eta
\end{equation}\\

Because equation (\ref{12}) is true for very strong shock waves, even if the
particle number is not conserved, relation (\ref{26}) remains valid in this case. If
we use as model an ultra-relativistic gas then equation (\ref{12}) is exact,
although it refers to $e_2$ instead of $W_2$. In addition, the variable $y$ becomes
\cite{IK} $$y=Ln(\frac{T}{T_0})$$ where $T$ is the absolute temperature of the gas.
Equation (\ref{26}) is replaced in this case by the following
\begin{equation}
\label{26a} y=2\frac{\nu-1}{\nu}\eta
\end{equation}
where $\nu$ is the arbitrary adiabatic index.\\

If in our problem (considering Bethe-Johnson gas) we assume that the particle
numbers are not conserved then for a very strong compressive shock ( $W\sim p/n>>1$)
thermodynamic calculations show that now $$y\approx Ln(\frac{T}{T_0})$$ and
\begin{equation}
\label{sigma} \frac{\sigma}{\sigma_0}\approx (\frac{T}{T_0})^{1/(\nu-1)}
\end{equation}
where $\sigma$ is the entropy per unit of volume  and we use $p\sim n^{\nu}$
 We can combine both expressions (\ref{26}) and (\ref{26a}) for $y$ into
one generalized relation
\begin{equation}
\label{26b} y=\delta\eta
\end{equation}
where $\delta$ can be either greater or less than $1$.\\

We substitute (\ref{26b}) in (\ref{25}), use (\ref{23}) and equation (\ref{24}) and
obtain ( after rather lengthy calculations) that at the shock ( where
$y=\delta\eta$)
\begin{equation}
\label{27} \{[(\delta +1)\frac{\partial}{\partial
y}+(\frac{\delta}{\nu-1}+1)\frac{\partial}{\partial\eta}](1-\frac{\partial}{\partial
y})\chi\}_{y=\delta\eta}=0
\end{equation}
The other boundary condition (at the Riemann wave) is borrowed from Ref.\cite{IK}
\begin{equation}
\label{28} \chi|_{\eta=y/\sqrt{\nu-1}}=0
\end{equation}
By introducing new variables
$$q=\eta-\frac{y}{\delta},~~~~~~~~~~~~~~~~~h=\frac{y}{\sqrt{(\nu-1)}}-\eta$$
we rewrite equation (\ref{24})and the boundary conditions (\ref{27}),(\ref{28})
\begin{equation}
\label{29} [(\frac{\nu-1}{\delta^2}-1)\frac{\partial^2}{\partial
q^2}+2(1-\frac{\sqrt{\nu-1}}{\delta})\frac{\partial^2}{\partial q\partial
h}+\frac{2-\nu}{\sqrt{\nu-1}}\frac{\partial}{\partial
h}-\frac{2-\nu}{\delta}\frac{\partial}{\partial q}]\chi=0
\end{equation}
\begin{equation}
\label{30}
\{[(\frac{\delta+1}{\sqrt{\nu-1}}-\frac{\delta+\nu-1}{\nu-1})\frac{\partial}{\partial
h}+\frac{\delta^2-nu+1}{\delta(\nu-1)}\frac{\partial}{\partial
q}](1+\frac{1}{\delta}\frac{\partial}{\partial q}-\frac{1}{\sqrt{\nu-1}}
\frac{\partial}{\partial h}\}\chi|_{q=0}=0
\end{equation}
Applying Laplace transform to variable $h$ in equations (\ref{28})-(\ref{30}) we
obtain
\begin{equation}
\label{31}
(\frac{\nu-1}{\\delta^2}-1)\frac{d^2\chi}{dq^2}+[2(1-\frac{\sqrt{\nu-1}}{\delta})r-
\frac{2-\nu}{\delta}]\frac{d\chi}{dq}+\frac{2-\nu}{\sqrt{\nu-1}}r\chi=0
\end{equation}
and
\begin{equation}
\label{32} [(\frac{\delta+1}{\sqrt{\nu-1}}-\frac{\delta+\nu-1}{\nu-1})r
+\frac{\delta^2-nu+1}{\delta(\nu-1)}\frac{d}{dq}](1+\frac{1}{\delta}\frac{d}{dq}-
\frac{1}{\sqrt{\nu-1}}r)\chi|_{q=0}=0
\end{equation}
 where $r$ is the transformation parameter.\\

 We represent a solution to (\ref{31}), (\ref{32}) as $a(r)e^{qk(r)}$. This leads
 to $2$ algebraic equations:
 \begin{equation}
 \label{33}
 (1-\hat{\delta}^2)\hat{k}^2+\hat{k}[2(\hat{\delta}-1)
 \hat{r}-\frac{2-\nu}{\nu-1}]+\frac{2-\nu}{\nu-1}\hat{r}=0
 \end{equation}
\begin{equation}
\label{34}
(\hat{k}+\frac{\sqrt{\nu-1}-1}{\hat{\delta}+1}\hat{r})(1+\hat{k}+\hat{r})a+
\frac{1}{\sqrt{\nu-1}}\frac{\sqrt{\nu-1}-1}{\hat{\delta}+1}\hat{\lambda}=0
\end{equation}
where $$\hat{k}=\frac{k}{\delta},~~~~\hat{r}=\frac{r}{\sqrt{\nu-1}},~~~~~
\hat{\delta}=\frac{\delta}{\sqrt{\nu-1}},~~~~~~\hat{\lambda}=\frac{\lambda}{\sqrt{\nu-1}}$$
In turn, the variable $\lambda$ is calculated at the point where Riemann's wave
catches up with the shock wave \cite{SB}
$$\lambda=(\frac{\partial\chi}{\partial q})|_{q=0,h=0}=
-(\frac{\partial\chi}{\partial h})|_{q=0,h=0}$$ \\

We will be interested in finding the entropy change in this flow. It is given by the
following expression \cite{SB}
\begin{equation}
\label{35} \Delta S=\Sigma_0\int_{t1}^{t_2}\sigma u_2dt
\end{equation}
Here $\Sigma_0$ is the "tunnel's" cross-sectional area, $\sigma$ is the entropy
density behind the shock, $u_2=\beta_2/\sqrt{1-\beta_2^2}$. $t_1$ is the moment of
time when Riemann's wave catches up with the shock wave and $t_2$ is the moment of
time when the shock wave reaches the right end of the "tunnel" (Fig.5). Since n we
consider a strong shock $\beta_2\rightarrow \nu-1$, that is
$u_2\approx(\nu-1)/\sqrt{2-\nu}$. Because the time interval  $dt$ is measured in the
shock   frame reference the respective time interval in the  observer frame ( where
initially both waves, the strong shock and Riemann's wave move with a velocity $D$)
is $dt/\sqrt{1-D^2}$. Taking these facts into account and using
(\ref{add}),(\ref{25}),(\ref{26b}) and (\ref{sigma}) we obtain from (\ref{35})
\begin{equation}
\label{36} \Delta S
=\frac{\delta^2-\nu+1}{\delta(\delta+1)}\int_0^{y_c}e^{y(2-\nu)/(\nu-1)}
[\frac{\partial}{\partial\eta}(\frac{\partial}{\partial y}-1)\chi]|_{y=\delta\eta}
\end{equation}

where $y_c$ is the value of $y$ at the moment when the shock wave reaches the right
end of the "tunnel".\\

By introducing variables $q$ and $h$ into (\ref{36}) and effecting the integration
we find after rather involved calculations
\begin{eqnarray}
\label{37} \Delta S=a_1([1- h'e^{a_2}\pm
a_3e^{h'A}\int_0^{h'_c}\frac{e^{-Az}-e^{-zF_{\pm}}}{h_c'-z}Z_{\pm}(h_c'-z)dz]
\end{eqnarray}
where
$$a_1=\frac{\pm\Sigma_0\sigma_0\lambda}{2}\frac{(A+\sqrt{A^2\mp 1})(1+\sqrt{\nu-1})}
{A+\sqrt{A^2\mp 1}+A\sqrt{\nu-1}},$$ $$a_2=-2A\frac{\nu-1}{2-\nu}
(1+\sqrt{\frac{1\mp A^{-2}}{\nu-1}}),$$
$$a_3=\frac{(A-\sqrt{A^2\mp 1})(\nu-2\sqrt{\nu-1}}{2-\nu},$$
\[ A^2=\left \{ \begin{array}{cc}
\hat{\delta}^2/(1-\hat{\delta}^2) & \mbox{for $\hat{\delta}<1$}\\
\hat{\delta}^2/(\hat{\delta}^2- 1) & \mbox{for $\hat{\delta}>1,$}
 \end{array}
 \right.\]
 $$h'=\frac{h}{2\sqrt{\nu-1}}\frac{2-\nu}{A-\sqrt{A^2\mp 1}},$$
 $$F_{\pm}=\frac{\nu A+2\sqrt{(A^2\mp 1)(\nu-1)}}{2-\nu},$$
\[ Z_{\pm}=\left\{ \begin{array}{ll}
I_1(h_c'-z) & \mbox{Bessel function of the imaginary argument}\\
J_a(h_c'-z) & \mbox{Bessel function of the first kind},
\end{array}
\right.\] and everywhere  the upper and lower signs correspond to $A>1$ and $A<1$
respectively.\\

We also have to find the value of $h_c$ entering integral in (\ref{37}). To this end
we use the relation from Ref.\cite{GM}
\begin{equation}
\label{38} x_c-l=t_0-t_c
\end{equation}
where $t_0=\lambda/\sqrt{\nu-1}$ is the moment of time when Riemann's wave overtake
the shock wave ( in the observer's frame), L is the coordinate corresponding to this
location of the wave, $x_c$ and $t_c$ are the coordinate and moment of time
describing the state of the shock weave at the right end of the "tunnel". Inserting
$x$ and $t$ from (\ref{23}) in (\ref{38}) and performing rather lengthy
calculations, we obtain
\begin{eqnarray}
\label{39}
\frac{(L\sqrt{\nu-1}+\lambda)(\sqrt{\nu-1}-\nu+1)}{\lambda\sqrt{\nu-1}}=\nonumber\\
\frac{2\sqrt{\nu-1}+\nu}{2A}\{A+\sqrt{A^2\mp 1}exp[-2\nu
Ah'/(2-\nu)]+A-\sqrt{A^2\mp 1}\}-\nonumber\\
\frac{2-\nu}{2A}e^{-g_{\pm}h'}\int_0^{h_c'}\frac{e^{g_{\pm}z}-e^{f_{\pm}z}}{h_c'-z}
Z_{\pm}(h_c'-z)dz
\end{eqnarray}where
$$g_{\pm}=\frac{\nu A-2\sqrt{(A^2\mp 1)(\nu-1)}}{2-\nu},$$
$$f_{\pm}=-[\nu A+2\sqrt{(A^2\mp 1)(\nu-1)}]$$
In what follows we neglect the integral in (\ref{39}) which would
introduce an error not exceeding $2\%$.\\

From the kinematic considerations follow that
$$L+\frac{\lambda}{\sqrt{\nu-1}}=(l+d)+\frac{l-d}{\nu-1}$$ and
$$\lambda=d(\frac{D+\beta_s}{D-\beta_s}+1)$$
Therefore relation (\ref{39}) (with the integral dropped) would yield the expression
for $exp(-2Ah'/\nu )$ in terms of $\nu$ and $A$
\begin{eqnarray}
\label{40} \pm(A+\sqrt{A^2\mp 1})^2exp(-\frac{2A\nu
h'}{2-\nu})=\nonumber\\
\frac{1}{\sqrt{\nu-1}(A-\sqrt{A^2\mp 1})}[\nu
A(1-\sqrt{\nu-1})\frac{l}{d}+\sqrt{(A^2\mp 1)(\nu-1)}-A(\nu-\sqrt{\nu-1})]
\end{eqnarray}\\

$i$) If $\delta=2(\nu-1)/\nu$ ( as in the case of an ultra-relativistic gas with an
arbitrary constant adiabatic index $\nu$) implying $A=2\sqrt{\nu-1}/(2-\nu)$ then we
choose in (\ref{37}) the lower sign. \footnote{ We could have chosen the upper sign.
However, the requirement $A>1$ would have left us with a very restricted range of
parameter $\nu$:$1\le\nu\le 4(1-\sqrt{1/2}$.} As a result, we obtain from (
\ref{37}) with the help of (\ref{40})
\begin{eqnarray}
\label{41} \frac{\Delta
S}{\Sigma_0\lambda\sigma_0}=\frac{3\nu-2+(\nu+2)\sqrt{\nu-1}}{2(3\nu-2)}
\{[(\frac{2-\nu}{\nu+2\sqrt{\nu-1}})^2(\frac{2\nu}{2-\nu}\frac{l}{d}-1)]^
{(3\nu-2)/2\nu}-1\}\nonumber\\
\end{eqnarray}
where $$\frac{l}{d}\ge\frac{1+\sqrt{\nu-1}}{1-\sqrt{\nu-1}}$$ The respective graph
of $<\Delta S>\equiv \Delta S/(\Sigma_0\lambda\sigma_0)$ is shown in Fig.5.
\begin{figure}[h]
\begin{center}
\includegraphics[width=6cm, height=6cm]{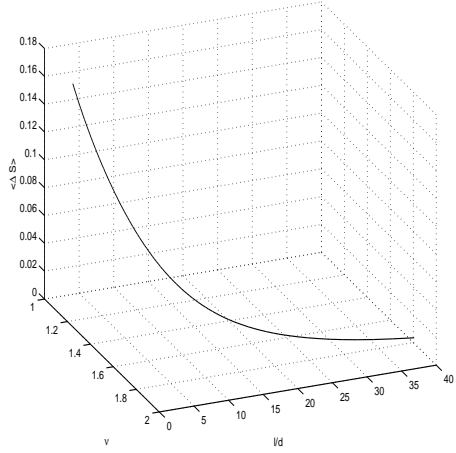}
\caption{\small {Variation of the dimensionless entropy change as a function of the
adiabatic index $\nu$ and $l/d$}}
\end{center}
\end{figure}
Expression given by (\ref{41}) yields as a particular case the expression found in
\cite{SB} for $\nu=4/3$ It is seen that the entropy change in region $2$ increases
and reaches its maximum for the values of $\nu$ in the range from $1.3$ to  $1.4$. A
further increase of $\nu$ results in a sharp decline of the entropy change,
approaching the purely adiabatic regime for the threshold value of $\nu=2$.
\\
$ii)$ If, we consider $\delta=2$ ( as in \ref{26}) we obtain in the same fashion as
in $i$:
\begin{equation}
\label{42} \frac{\Delta S}{\Sigma_0\lambda\sigma_0}=\frac{3\sqrt{\nu-1}+\nu+1}
{6\sqrt{\nu-1}}\{1-[1+\frac{2\nu}{\nu-1+2\sqrt{\nu-1}}
(\frac{l/d}{l_c/d}-1)]^{3(\nu-1)/2\nu}\}
\end{equation}
where $l_m/d$ is given by (\ref{21}). From (\ref{42}) follows that in this case the
entropy change can be positive only if $\nu>2.$ However this also means that the
flow becomes non-causal , since ( as we showed above) in this case $\beta_s>1.$
Therefore if $\delta =2$ the one-dimensional flow of Bethe-Johnson gas with $\nu<2$
is characterized by a zero change of entropy when Riemann's wave catches up with the
shock wave (and stays zero from that moment on), that is the flow becomes adiabatic.
This contrasts with the previous case $i)$ which is a generalization of the
analogous flow for an ultra-relativistic with an arbitrary adiabatic index ( not
necessarily $\nu=4/3$ as in \cite{SB}). In the latter the entropy change is linear
in $l/d$ before Riemann wave catches up with the shock wave, and obeys the power law
for $l/d$ (given by eq.\ref{42}) after Riemann wave overtakes the shock wave. \\

The flow described here can be associated with a nucleon-nucleon collision
(cf.\cite{SB1}) which is modelled as a collision of  the nucleon with a cylinder cut
out of the nucleus and whose cross-sectional area is equal to the one of the
nucleus. The nucleus linear size varies from the diameter of the nucleus to the
"diameter" of the nucleon, $d$. It is clear that such an approximation can be used
only to obtain certain estimates about parameters of the real process, for example
to evaluate an entropy change. The ratio $l/d$ is related to the atomic number
$\bf{A}$: $l/d={\bf{A}}^{1/3}.$ If we take  $max{\bf{A}}=257$ this would give us the
maximum value of $\nu=1.53$ ( according to eq.\ref{21})which is close to the region
where entropy change reaches its maximum.
\section{CONCLUSION}
We found the shock adiabatic for BJ gas and applied the found detailed relations for
a study of a one-dimensional flow of such a gas in a context of a nucleon-nucleon
collision.  If baryon matter can be transformed into quark matter at a certain range
of the compression ratio ($5-10$), then the proposed model allows one to identify
the baryon matter as Bethe-Johnson gas with the adiabatic index $\nu\approx 1.5.$\

Such an approach generalizes an analogous approach based upon an ultra-relativistic
gas. The latter represents a particular case of BJ gas with adiabatic index
$\nu=4/3$. The entropy change behind the shock strongly depends on the choice of the
adiabatic index in BJ, by reaching its maximum in the range of values of $\nu$ from
$1.3$ to $1.4$.

\end{document}